\documentclass[sigconf, balance=false]{acmart}
\usepackage{popets}
\usepackage{algorithm}
\usepackage{algpseudocode}
\usepackage{bbm}

\acmYear{YYYY}
\acmVolume{YYYY}
\acmNumber{X}
\acmDOI{XXXXXXX.XXXXXXX}
\acmISBN{}
\acmConference{Proceedings on Privacy Enhancing Technologies}
\usepackage[english]{babel}
\newtheorem{assumption}{Assumption}
\newcommand{\indep}{\perp \!\!\! \perp}
\begin{document}

\title[LOCKS]{LOCKS: User Differentially Private and Federated Optimal Client Sampling}

\author{Ajinkya K Mulay}
\orcid{1234-5678-9012}
\affiliation{%
  \institution{Purdue University}
  \city{West Lafayette}
  \state{IN}
  \country{USA}}
\email{mulay@purdue.edu}


\renewcommand{\shortauthors}{Mulay et al.}

\begin{abstract}
With changes in privacy laws, there is often a hard requirement for client data to remain on the device rather than being sent to the server. Therefore, most processing happens on the device, and only an altered element is sent to the server. Such mechanisms are developed by leveraging differential privacy and federated learning. Differential privacy adds noise to the client outputs and thus deteriorates the quality of each iteration. This distributed setting adds a layer of complexity and additional communication and performance overhead. These costs are additive per round, so we need to reduce the number of iterations. In this work, we provide an analytical framework for studying the convergence guarantees of gradient-based distributed algorithms. We show that our private algorithm minimizes the expected gradient variance in $\approx d^2$ rounds, where $d$ is the dimensionality of the model. We discuss and suggest novel ways to improve the convergence rate to minimize the overhead using Importance  Sampling (IS) and gradient diversity. Finally, we provide alternative frameworks that might be better suited to exploit client sampling techniques like IS and gradient diversity.
\end{abstract}
\keywords{local differential privacy, high-dimensional linear models, federated learning}

\maketitle


\section{Introduction}

Stochastic Gradient Descent (SGD) functions by randomly selecting a subset of data points per round and computing their gradients. The update step in SGD uses the average gradient to modify the weights from the previous round. Optimal model weights can improve the trained model and its inference capabilities on unseen data. For high-dimensional non-convex parameter paces, obtaining optimal models is non-trivial and computationally heavy. Thus, improving the quality of gradients and accelerating the training process are fundamental goals for improving gradient algorithms.   

However, as we often see in practice, the information assimilated in each gradient can vary widely. After the first few training epochs, the gradients of specific data points lose their value. \citep{katharopoulos2018not}. We could select data points randomly or by batching the entire training set per round. Both methods are inefficient and do not consider the information contributed by the gradients or the data points.

A way to assess the importance of each data point could be based on the magnitude of the gradient or loss value contributed by the data point \citep{loshchilov2015online}, \cite{alain2015variance}. Gradient magnitude refers to metrics like the $L_1$-norm or $L_2$-norm computed on individual gradients. Larger magnitudes suggest higher levels of information in gradients. For high-dimensional parameter models, gradient norms can be costly to compute. Computing loss values is straightforward but needs to reflect the gradient update accurately. In this work, we, therefore, focus on the magnitude of the gradient update. Such methods are based on Importance Sampling (IS) algorithms.

The rationale behind randomly picking data points per round is to obtain an unbiased estimate of the actual gradient. However, this unbiasedness comes at the cost of higher gradient variance. In such cases, Stochastic Variance Reduction Gradient (SVRG) techniques have been suggested to reduce the variance of the gradient.

Besides the gradient magnitude, we should prioritize diverse gradients. If two gradients are similar in the form of some similarity metric (ex., cosine similarity), then we may only need to use such a gradient. Gradient diversity should therefore be another basis for client selection. 

This work further focuses on distributed and private machine-learning settings. Big data technologies have enabled massive leaps in predictive learning. However, these techniques commonly access private personal client data that can potentially cause severe harm to the owner if leaked. Such leaks are preventable under Federated Learning (FL) and Differential Privacy (DP), allowing owners tighter ownership over their data. For vanilla SGD, there are already techniques like FedAvg \citep{konevcny2016federated} and DP-SGD \citep{abadi2016deep} that can partially or fully protect private data. However, these algorithms' privacy and convergence analysis cannot be directly applied to the algorithms proposed in this work. 

In this work, we study the problem of partial client availability under Federated Learning under differential privacy. We analyze the expected gradient variance due to uniform sampling and the noise added by DP with adaptive server learning rates. As per our knowledge, such an exact theoretical analysis does not exist. Therefore, we provide analytical results and convergence guarantees in Theorem~\ref{theorem:main1} in terms of the required iteration. We also suggested a modified approach that allows for lower communication rounds by simple scaling. We provide a pseudo-code for the algorithm suggested. We motivate the identification of other frameworks due to significant drawbacks to the one suggested in the reference papers.

\section{Motivation}

Federated Learning involves training machine learning models on individual clients rather than sending the entire dataset to a central server. Clients process and train their datasets with an SGD variant on-device. 
At each round, the server selects a subset of its client for training. Clients only send their computed gradient to the server. The server computes the average gradient over these client gradients and updates the model. After each round, the server sends back the updated model. These rounds are repeated until the model converges, similar to vanilla training. With more rounds, the communication cost incurred by the clients and the server substantially increases. Also, with limited computational resources and data, the clients must train longer and for more rounds. Therefore, we prefer to lower the number of training rounds.

Differential Privacy (DP) is a mathematical framework over queries to a dataset that limits the leakage of a single client or data point in a dataset. However, even though we wish to learn aggregated values, the query outputs can disclose information about a particular individual in the dataset. Combining auxiliary public data with aggregated histograms or anonymized datasets can help us extract exact individual identities \citep{narayanan2006break}. DP adds noise or performs specific sampling to limit the impact of a single client or data point on the query outputs, thus protecting individual privacy. When accessing data over multiple rounds, DP needs to add noise at each stage; thus, the DP cost is proportional to the number of rounds. Higher DP costs lead to practically trivial privacy for real-world applications. Thus, reducing the number of rounds is crucial to ensuring high privacy standards.

Therefore, an accelerated SGD algorithm will improve training quality and speeds and potentially lead to more private methods. Federated IS (FedIS) and Differentially Private IS (DPIS) leverage gradient magnitude and diversity to achieve this acceleration. We study approaches that combine some of these ideas and introduce new ones for private training.

\section{Preliminaries}

\subsection{Differential Privacy}

DP proposes a solution to publishing or sharing private datasets. DP mathematically defines privacy leakage caused due to computations over private data points. By providing an attacker-agnostic and information-agnostic defense, DP protects against worst-case scenarios. Thus, we do not need specific assumptions about the attacker and are protected against future information releases.

In the DP literature, we commonly consider a dataset as a multi-set of $n$ samples in the $\mathcal{X}^n$ space. This is analogous to the dataset belonging to the real space $\mathbb{R}^{n \times p}$. The following section is based on preliminary definitions found in \citep{dwork2014algorithmic}.

To understand the mathematical notion of DP, we define pure and approximate DP. 

\begin{definition}(Differential Privacy) A randomized mechanism $\mathcal{M}: \mathcal{X}^n \rightarrow \mathbb{R}^{d}$ is considered to be $(\epsilon, \delta)$ DP if for all neighboring datasets $x, x^{'} \in \mathcal{X}^n$ differing in a single sample (\emph{i.e., } $||x-x^{'}|| \leq 1$), we have
\begin{align}
    \text{Pr}[\mathcal{M}(x) \in \mathbb{R}^{d}] \leq e^{\epsilon}\text{Pr}[\mathcal{M}(x^{'}) \in \mathbb{R}^{d}] + \delta
\end{align}
    
\end{definition}

Differential privacy allows for pure-DP ($\delta=0$) or approximate-DP ($\delta>0$), each delivered by a different noise addition mechanism. Approximate DP allows for an infinitesimally small probability of data leakage. However, it composes (adds up) well over multiple iterations compared to pure DP. For most computations in this work, we will stick to approximate DP. 

\begin{definition} ($l_k$-Sensitivity) For a function $f: \mathcal{X}^n \rightarrow \mathbb{R}^d$, we define its $l_k$ norm sensitivity (denoted as $\Delta_k f$) over all neighboring datasets $x, x^{'} \in \mathcal{X}^n$ differing in a single sample as 
\begin{align}
    \text{sup}_{x, x^{'} \in \mathcal{X}^n} ||f(x) - f(x^{'})||_k \leq \Delta_k f 
\end{align}
\end{definition}

Given the $l_2$ sensitivity of a function, we can further define the Gaussian Mechanism that adds noise to the function $f: \mathcal{X}^n \rightarrow \mathbb{R}^d$. Gaussian DP is used to achieve approximate DP.

\begin{lemma}(Gaussian Mechanism)
    Consider an arbitrary $\epsilon \in (0, 1)$. Let the $l_2$ sensitivity of the function $f: \mathcal{X}^n \rightarrow \mathbb{R}^d$ be $\Delta_2 f$. Further, consider $\sigma = \frac{c\Delta_2f}{\epsilon}$ where $c^2 > 2ln(\frac{1.25}{\delta})$. Then, the following randomized mechanism $\mathcal{M}$,
    \begin{align}
        \eta &\sim \mathcal{N}(0, \sigma^2) \\
        \mathcal{M}(x, \sigma) &\coloneqq f(x) + \eta
    \end{align}
    satisfies approximate or $(\epsilon, \delta)$ DP. 
\end{lemma}

We define a few other properties of DP. 

\begin{definition}(Rényi Divergence) We use Rényi divergence to quantify the closeness of two probability distributions. Given, two probability distributions $\mathcal{P}_1, \mathcal{P}_2$ and their density functions $p_1(x), p_2(x)$ s.t. $x \in \mathcal{X}$ then the Rényi divergence of order ($\alpha > 1$) is given as,
\begin{align}
    D_{\alpha}(\mathcal{P}_1 || \mathcal{P}_2) \coloneqq \frac{1}{\alpha - 1} ln \int_{\mathcal{X}} q(x) (\frac{p(x)}{q(c)})^{\alpha} dx 
\end{align}
\end{definition}

Based on the Rényi divergence, we use the Rényi Differential Privacy (RDP) \citep{mironov2017renyi} to quantify our proposed algorithms' privacy cost. Although approximate DP is a stricter notion of privacy, its composition analysis is weaker than the newer forms of privacy like RDP, Concentrated Differential Privacy (CDP), and z-CDP. In this case, we choose to use RDP to be consistent with the privacy costs shown in \citep{wei2022dpis}.

\begin{definition}(Rényi Differential Privacy (RDP)) For any neighboring datasets $x, x^{'} \in \mathcal{X}^n$, a randomized mechanism $\mathcal{M}: \mathcal{X}^n \rightarrow \mathbb{R}^{d}$ satisfies $(\alpha, \tau)$-RDP if
\begin{align}
    D_{\alpha}(\mathcal{M}(x) || \mathcal{M}(x^{'})) \leq \tau
\end{align}    
\end{definition}

\begin{definition}($(\alpha, \tau)$-RDP to $(\epsilon, \delta)$-DP conversion) The randomized mechanism $\mathcal{M}: \mathcal{X}^n \rightarrow \mathbb{R}^{d}$, satisfies $(\alpha, \tau)$-RDP. For the constants, $\alpha > 1, \epsilon \geq 0$ and appropriate $\delta, \mathcal{M}$ satisfies $(\epsilon, \delta)$-DP for,
\begin{align}
    \epsilon = \tau + \frac{\text{log}(\frac{1}{\delta}) + (\alpha - 1)\text{log}(1 - \frac{1}{\alpha}) - \text{log}(\alpha)}{\alpha - 1}
\end{align}
    
\end{definition}

\subsection{Federated Optimization}

Federated Learning pursues distributed optimization by allowing clients to train models on their data. The raw data never leaves the device, and each client only sends the gradient value to be aggregated on the server. Thus, the optimization function is split across $n$ clients and can be arranged as the following,
\begin{align}
    w^{*} &\coloneqq \text{arg min}_{w \in \mathbb{R}^p} F(w) 
    \\ & \coloneqq \text{arg min}_{w \in \mathbb{R}^p} \frac{1}{n} \sum_{i=1}^{n} f(w; X_i, y_i) 
    \\ & \coloneqq \text{arg min}_{w \in \mathbb{R}^p} \frac{1}{n} \sum_{i=1}^{n} f_i(w) 
\end{align}
Here, $X$ and $y$ are the dataset and response variable resp.

We assume a non-convex setting and make the following assumptions as made in \citep{Delta:FasterFederatedLearning} (restated for completeness). 

\subsubsection{Assumptions}

\begin{assumption}[L-smooth]\label{assumption:lsmooth} Each client has a local objective function that is $L$-Lipschitz Smooth ($L > 0$) s.t. 
\begin{align}
    ||\triangledown f_i(w_1) - \triangledown f_i(w_2)||_2 \leq L ||w_1-w_2||_2, \forall w_1, w_2 \in \mathbb{R}^p; i \in [n] 
\end{align}
\end{assumption}

\begin{assumption}(Unbiased Local Gradient Estimator and Local Variance)\label{assumption:unbiased} Suppose for the $t^{th}$ round we have sample $i, x_{t}^{i}$ and model parameter $w_t$. Then for all clients $i \in [n]$ 
\begin{align}
E_{x_{t}^{i} \sim \mathbb{U}(S^i)}[\triangledown f_i(w, x_{t}^{i})] = \triangledown f_i(w)
\end{align} where $S^i$ is the set of all the data samples at client $i$ and $\mathbb{U}$ is the uniform distribution.

Further, the local utility functions have bounded local variance s.t., 
\begin{align}
    E_{x_{t}^{i} \sim \mathbb{U}(S^i)}[||\triangledown f_i(w, x_{t}^{i}) -  \triangledown f_i(w)||^2] = \sigma_{L, i}^2 \leq \sigma_L^2
\end{align}
    
\end{assumption}

\begin{assumption}(Bound Dissimilarity)\label{assumption:bounded-dissimilarity} \cite{wang2020tackling} The following expression relates the expected value of the local and global objective functions,
\begin{align}
    \mathbb{E}_t||\triangledown f_i(w)||^2_2 \leq (A^2 + 1)||\triangledown F(w)||^2_2 + \sigma^2_G
\end{align}
where $A, \sigma_G \leq 0$ are constants. For the case when $A = \sigma_G = 0$, then all local loss functions are similar.
\end{assumption}

\begin{assumption}(Bounded Clipping) \label{assumption:clipped}
Suppose vector $x$ is clipped s.t. $x_c = \frac{x}{\max \{1, ||x||_2/C\}}$ then we define $\nu_{x} \coloneqq x_c - x_c$. We say that 
\begin{align}
    E[||\nu_{x}||^2] \leq c_{\nu}^2
\end{align}
\end{assumption}

The last assumption is a new one that suggests that the variance due to clipping is generally small. \citep{Delta:FasterFederatedLearning} 
also assumes that the stochastic gradient's norm is uniformly bounded. However, since we will be clipping the local gradients before adding noise for DP, we do not require this explicit assumption. Rather, we use the clipping constant to bound each local gradient's norm.

\section{Preliminary Work}

\begin{lemma}\label{lemma:expected-batchsize}
    The expected batch size ($\mathbb{E}[|\mathcal{Z}_{m, t}|]$) due to Algorithm~\ref{algorithm:LOCks} is less than b.  Here $n$ is the number of total clients, $m$ is the fraction of clients chosen in the first round, and $\frac{m}{k} = b$ where $n, k, m \in \mathbb{Z}^{+}$. Further note that $p_{i, t} \leq 1$ for all $i \in [n]$. The values are defined as shown in Algorithm~\ref{algorithm:LOCks}.
\end{lemma}
\textbf{Proof:} We know that $\mathbb{E}[|\mathcal{Z}_{m, t}|]$ can be written as the follows,
\begin{align}
    \mathbb{E}[|\mathcal{Z}_{m, t}|] &= \sum_{i=1}^n 1 \cdot \frac{m}{n} 
    \frac{p_{i, t}}{k}
    \\ & = \sum_{i=1}^n \frac{b}{n} p_{i, t}
    \\ & \leq \sum_{i=1}^n \frac{b}{n} = b
\end{align}

\begin{lemma}\label{lemma:unbiased-estimator}
    The private gradient estimate at round $t$, ($\tilde{\Delta}_t$) in Algorithm~\ref{algorithm:LOCks} is an unbiased estimator of the aggregate clipped gradients.
\end{lemma}
Let us first understand the expression for $\tilde{\Delta}_t$. Note that $\bar{u}_t = \frac{1}{b}\sum_{l \in \mathcal{Z}_{m, t} u_l}$ and thus by definition has expected value of zero. (1) follows from this idea. (2) follows from introducing the next sampling probability of clients in the first and second rounds (combined).
\begin{align}
    \mathbb{E}_t[\tilde{\Delta}_t] &= \mathbb{E}_t[\Delta_t + \bar{u}_t]
    \\ & \overset{(1)}{=} \mathbb{E}_t[\Delta_t]
    \\ & = \mathbb{E}_t[\frac{1}{b} \sum_{j \in \mathcal{Z}_{m. t}}\frac{\bar{g}_j(w_t)}{p _{j, t}}]
    \\ &= \mathbb{E}_t[\frac{1}{b} \mathbbm{1}[i \in \mathcal{Z}_{m, t}] \sum_{i = 1}^n \frac{\bar{g}_i(w_t)}{p_{i, t}}]
    \\ &= \sum_{i=1}^n \bar{g}_i(w_t) \frac{1}{b p_{i, t}} \mathbb{E}_t[\mathbbm{1}[i \in \mathcal{Z}_{m, t}]]
    \\ & \overset{(2)}{=} \sum_{i = 1}^n \bar{g}_i(w_t) \frac{1}{bp _{i, t}} \frac{q_{i, t} p_{i, t}}{k}
    \\ & = \sum_{i = 1}^n \bar{g}_i(w_t) \frac{1}{b} \frac{m}{nk}
    \\ & = \frac{1}{n}\sum_{i = 1}^n \bar{g}_i(w_t) = \bar{\Delta}_t
\end{align}

\section{Main Theorems}
We need to find an optimal sampling mechanism for our proposed method that minimizes overall variance. Additional variance is introduced due to the partial participation of clients leading to an approximation of the overall objective. We first study the impact of uniformly choosing samples from the set in Theorem~\ref{theorem:main1}.


\begin{theorem}\label{theorem:main1}
    Let the adaptive global learning rates be $\eta_{t}$. Under our assumptions~\ref{assumption:lsmooth}-\ref{assumption:clipped}, the model parameter sequence $\{w_t\}$ generated by Algorithm~\ref{algorithm:LOCks} satisfies the following:
    \begin{align}
        \min_{t \in [T]} \mathbb{E}[||\triangledown F(w_t)||_2^2] &\leq \frac{F_(w_0) - F_(w_{*})}{\sqrt{T}\alpha_{\kappa}} + \Phi
        \\ & = \mathcal{O}(\frac{\Delta F}{\sqrt{T}} + \sigma_L^2 + c_{\nu}^2 + \sigma_0^2 + \frac{\sigma^2_G}{d})
    \end{align}
    where the expectation is taken over the local dataset samples among clients. $\Phi$ is the variance seen due to random sampling and privacy noise. Here, $T = d^2 log(T)^2$. We uniformly sample clients as shown in Algorithm~\ref{algorithm:LOCks} with a net probability $\frac{bV_0}{n}$ (across both rounds of sampling). The variance bounds $\sigma_L, \sigma_G, c_{\nu}^2$ are introduced in assumptions~\ref{assumption:lsmooth}-\ref{assumption:clipped}. $d$ is the number of model parameters, $\sigma_0$ is the privacy noise added as per Algorithm~\ref{algorithm:LOCks}. Furthermore,
    \begin{align}
        \Delta F &= F(w_0) - F(w_{*}) \\
        \alpha_{\kappa} &= \eta_b \frac{\sqrt{T} - 1}{\sqrt{T}} - \frac{V_0L\eta_b^2 log(T)(A^2+1)}{b\sqrt{T}} \\
        \\ \Phi &= \frac{(2 \eta_b K_{max}(\sigma^2_L + c_{\nu}^2))}{K_{min}\alpha_{\kappa}} 
        \\ & + \frac{L\eta_b^2}{b\alpha_{\kappa}\sqrt{T}}[\frac{K_{max}(\sigma_L^2 + c_{\nu}^2)}{K_{min}} + \frac{\sigma^2_G + d\sigma^2_0}{2}] V_{0} log(T)
    \end{align}
\end{theorem}
\textbf{Proof:} We define $\bar{\Delta}_t \coloneqq \frac{1}{n} \sum_{i=1}^n \tilde{g}_i (w_{t})$. In the case all participants participate, we can see that $\bar{\Delta}_t = \tilde{\Delta}_t$. However, for partial client participation we see that $\bar{\Delta}_t = \sum_{i=1}^n \tilde{g}_i (w_{t}) \neq \sum_{i \in \mathcal{i \in \mathcal{X}_{m, t}}} \tilde{g}_i (w_{t}) =\tilde{\Delta}_t$. Thus, the randomness in partial client participation consists of client sampling, stochastic gradient, and privacy noise. 

Due to the smoothness assumption on $F$ (by assumption ~\ref{assumption:lsmooth}), and taking the expectation of $F(w_{t+1})$ over the client sampling and at round $t$. Note $w_t$ is unaffected by the sampling at round $t$.
\begin{align}
    \mathbb{E}&_t[F(w_{t+1})] \nonumber
    \\ &\leq F(w_t) + \langle \triangledown F(w_t), \mathbb{E}_t[w_{t+1} - w_{t}] \rangle + \frac{L}{2} \mathbb{E}_{t}[||w_{t+1} - w_t||^2] \nonumber
    \\ & = F(w_t) + \langle \triangledown F(w_t), \eta \mathbb{E}_t[\tilde{\Delta}_t]\rangle + \frac{L}{2}\eta^2 \mathbb{E}_t[||\tilde{\Delta}_t||^2] \nonumber
    \\ & = F(w_t) + \langle \triangledown F(w_t), \eta \mathbb{E}_t[\Delta_t + \bar{u}_{t}] \rangle + \frac{L}{2}\eta^2 \mathbb{E}_t[||\Delta_t + \bar{u}_{t}||^2] \nonumber
\end{align}
Here, $\bar{u}_{t} = \frac{1}{|\mathcal{Z}_{m, t}|}\sum_{i \in \mathcal{Z}_{m,t}} u_i$ and then $\bar{u}_t \sim \mathcal{N}(0, \sqrt{|\mathcal{Z}_{m, t}|}\sigma_0^2)$. Further, $\Delta_t = \sum_{i \in \mathcal{Z}_{m, t}} \frac{\bar{g}_i(w_t)}{n q_{i, t} p_{i, t}}$. We note that expectation over the randomness of client sampling and private Gaussian noise gives us $\mathbb{E}_{t}[\Delta_t + \bar{u}_t] = \sum_{i \in \mathcal{Z}_{m, t}}\mathbb{E}_{t}[\frac{\bar{g}_i(w_t)}{bn q_{i, t} p_{i, t}} + \bar{u}_t] = \sum_{i \in \mathcal{Z}_{m, t}} \mathbb{E}_{t}[\frac{\bar{g}_i(w_t)}{bn q_{i, t} p_{i, t}}] = \mathbb{E}_{t} [\Delta_t]$. Due to the independent Gaussian noise and expectation over its randomness, we have $\mathbb{E}_t[\bar{u}_t] = 0$ and $\Delta_t \indep \bar{u}_t \rightarrow \mathbb{E}_t[\langle \Delta_t, \bar{u}_t \rangle] = 0$. 

Thus, $\mathbb{E}_t[||\Delta_t + \bar{u}_t||^2_2] = \mathbb{E}_t[||\Delta_t||_2^2 + ||\bar{u}_t||^2_2 + 2\langle \Delta_t, \bar{u}_t \rangle] = \mathbb{E}_t[||\Delta_t||_2^2 + ||\bar{u}_t||^2_2]$ . Therefore,
\begin{align}
    \mathbb{E}&_t[F(w_{t+1})] \nonumber
    \\ & = F(w_t) - \eta_t ||\triangledown F(w_t)||^2_2 + \underbrace{\langle \triangledown F(w_t), \eta_t \mathbb{E}_t[\Delta_t + \triangledown F(w_t)]\rangle}_{A_1} \\ &+ \frac{L}{2}\eta_t^2 \underbrace{\mathbb{E}_t[||\Delta_t||^2]}_{A_2} + \underbrace{\frac{L}{2}\eta_t^2\mathbb{E}_t[||\bar{u}_t||^2]}_{A_3} \nonumber
\end{align}

Let us bound term $A_3$ first. We note that by our sampling technique, the expected batch size is $\mathbb{E}[|\mathcal{Z}_{m, t}|] \leq b$ (Lemma~\ref{lemma:expected-batchsize}) and that $\bar{u}_t \sim \mathcal{N}(0, \sqrt{|\mathcal{Z}_{m, t}|}\sigma_{0}^2\mathbb{I}_d)$.
\begin{align}
    A_3 &= \frac{L}{2}\eta_t^2\mathbb{E}_{t}[||\bar{u}_t||^2]
    \\ & \overset{(1)}{=} \frac{L}{2}\eta_t^2 \mathbb{E}_{t} [||\sum_{i=1}^n \mathbbm{1}[i \in \mathcal{Z}_{m, t}] \frac{u_i}{bp_{i, t}}||^2_2]
    \\ & =  \frac{L\eta_t^2}{2b^2} \sum_{i=1}^n \text{Pr}[i \in \mathcal{Z}_{m, t}]\mathbb{E}_{t} [||\frac{u_i}{p_{i, t}}||^2_2]
    \\ & = \sum_{i=1}^n 
    \frac{L\eta_t^2}{2b^2} \frac{p_{i, t} q_{i, t}}{kp_{i, t}^2} \mathbb{E}_{t} [||u_i||^2_2]
    \\ & = \sum_{i=1}^n \frac{L\eta_t^2}{2bn p_{i, t}} d\sigma_0^2
\end{align}
Here (1) is from the law of total expectation (or tower expression), (2) is due to $E[||x||_2^2] = E[\sum_{k=1}^p x_i^2] = p\sigma^2$ where $x_i \sim \mathcal{N}(0, \sigma^2)$, (3) is by Jensen's inequality for a concave function (squared root is concave) and the (4) follows from Lemma~\ref{lemma:expected-batchsize}.

Let us now bound $A_2$ with the assumption that we sample with replacement.
\begin{align}
    A_2 &= \mathbb{E}_t[||\Delta_t||_2^2]
    \\ &=\mathbb{E}_t[||\frac{1}{b} \sum_{j \in Z_{m, t}} \frac{\bar{g}_j(w_t)}{p_{j, t}}||^2]
    \\ & \overset{(1)}{=} \underbrace{\frac{1}{b^2} \mathbb{E}_t[||\sum_{j \in \mathcal{Z}_{m, t}} \sum_{r=1}^{K_{j}} \frac{\triangledown f_j(w_t; x_{jr}) - \triangledown f_j(w_t) - \nu_{jr}}{p_{j, t}K_j}||_2^2]}_{B_1}
    \\ & + \underbrace{\frac{1}{b^2} \mathbb{E}_t [||\sum_{j \in \mathcal{Z}_{m, t}} \frac{1}{p_{j, t}} \triangledown f_j(w_t)||_2^2]}_{B_2}
\end{align}
Here, (1) follows from the simple expression $E[||x||^2] = E[||x - E[x]||^2] + ||E[x]||^2$. We first bound the first term $B_1$. Note, $\nu_{jr}$ is the factor clipped off by clipping s.t. $\nu_{jr} \coloneqq \triangledown f_i(w_t, x_{jr} - \triangledown \bar{f}_i(w_t, x_{jr}$. Here, $\bar{f}_i(w_t, x_{jr}$ is the clipped form of $\triangledown f_i(w_t, x_{jr}$. We assume $E_t[||\nu_{jr}||_2^2] \leq c_{\nu}^2$.
\begin{align}
    B_1 &\overset{(1)}{=} \frac{1}{b^2}\mathbb{E}_t [||\sum_{i=1}^n \mathbbm{1}[i \in Z_{m, t}]\sum_{r=1}^{K_i}
    \frac{\triangledown f_i(w_t; x_{ir}) - \triangledown f_i(w_t) - \nu_{ir}}{p_{i, t}K_i}||^2] \nonumber
    \\ & = \frac{1}{b^2}||\sum_{i=1}^n \mathbb{E}_t[\mathbbm{1}[i \in Z_{m, t}]\sum_{r=1}^{K_i}
    \frac{\triangledown f_i(w_t; x_{ir}) - \triangledown f_i(w_t) - \nu_{ir}}{p_{i, t}K_i}]||^2 \nonumber
    \\ & \overset{(2)}{=} \sum_{i=1}^n \text{Pr}[i \in Z_{m, t}] \mathbb{E}_t [||  \frac{1}{bp_{i, t}K_i} \sum_{r=1}^{K_i} \triangledown f_i(w_t; x_{ir}) - \triangledown f_i(w_t) - \nu_{ir}||^2_2]  
    \nonumber
    \\ & = \sum_{i=1}^n \frac{p_{i, t}q_{i,t}}{k} \mathbb{E}_t [||\frac{1}{bp_{i, t}K_i} \sum_{r=1}^{K_i} [\triangledown f_i(w_t; x_{ir}) - \triangledown f_i(w_t) - \nu_{ir}]||^2_2 \nonumber
    \\ & = \sum_{i=1}^n \frac{p_{i, t}q_{i,t}}{kb^2 p_{i, t}^2 K_i^2} \mathbb{E}_t [||\frac{1}{bp_{i, t}K_i} \sum_{r=1}^{K_i} [\triangledown f_i(w_t; x_{ir}) - \triangledown f_i(w_t) - \nu_{ir}]||^2_2 
    \\ & \leq \sum_{i=1}^n \frac{1}{bn p_{i, t} K_{min}^2} \mathbb{E}_t [||\frac{1}{bp_{i, t}K_i} \sum_{r=1}^{K_i} [\triangledown f_i(w_t; x_{ir}) - \triangledown f_i(w_t) - \nu_{ir}]||^2_2 
    \\ &\overset{(3)}{\leq} \sum_{i=1}^n \frac{2K_{max}}{bnp_{i, t}K_{min}^2}(\sigma^2_L + c_{\nu}^2)
\end{align}
Here (1) rewrites the summation with the indicator variable trick, and in (2), we take the expected value of the indicator function (also equal to its probability). In (3), we use the inequality $E[||x_1 + x_2 + ... + x_n||^2] \leq n(E[||x_1||^2] + ... E[||x_n||^2)$ and $E[||x_1 - x_2||^2] \leq 2(E[||x_1||^2] + E[||x_2||^2)$, assumption~\ref{assumption:unbiased} and assume $K_{max} = \max_{i=1}^n K_i$ (achieved by clipping the number of clients per user) and $K_{min} = \min_{i=1}^n K_i$. The server broadcasts both $K_{max}$ and $K_{min}$. Clients limit their training to $K_{max}$ samples with repetition and ensure that at least $K_{min}$ samples (with replacement) are trained each time.

Let us bound $B_2$ now,
\begin{align}
    B_2 &= \frac{1}{b^2}  \mathbb{E}_t[||\sum_{i=1}^n \mathbbm{1}[i \in \mathcal{Z}_{m, t}] \frac{1}{p_{i, t}} f_i(w_t)||^2_2] 
    \\ & \overset{(1)}{=} \frac{1}{b^2}\sum_{i=1}^n \text{Pr}(i \in \mathcal{Z}_{m, t})E_{t}|| \triangledown f_i(w_t)||^2
    \\ & = \sum_{i=1}^n \frac{p_{i, t} q_{i, t}}{kb^2 p_{i, t}^2} E_{t}|| \triangledown f_i(w_t)||^2
    \\ & = \sum_{i=1}^n \frac{1}{bn p_{i, t}} E_{t}|| \triangledown f_i(w_t)||^2
    \\ & \overset{(2)}{\leq} \sum_{i=1}^n \frac{1}{bnp_{i, t}} ((A^2 + 1)||\triangledown F(w_t)||^2 + \sigma^2_G)
\end{align}
where (1) follows from analysis similar to $B_1$ and cancellation of probabilities, (2) uses the inequality $E[||x_1 + x_2 + ... + x_n||^2] \leq n(E[||x_1||^2] + ... E[||x_n||^2)$ and Assumption~\ref{assumption:bounded-dissimilarity}.

Now, we only need to bound $A_1$.

\begin{align}
    A_1 &= \langle \triangledown F(w_t), \eta_t \mathbb{E}_t[\Delta_t  +  \triangledown F(w_t)]\rangle
    \\ &= \langle \triangledown F(w_t), \eta_t \mathbb{E}_t [\sum_{i=1}^n \mathbbm{1}[i \in Z_{m, t}] \frac{-\bar{g}_i(w_t)}{bp_{i, t}} + \triangledown F(w_t)] \rangle
    \\ &=\langle \triangledown F(w_t), -\eta_t \mathbb{E}_t [\sum_{i=1}^n \mathbbm{1}[i \in Z_{m, t}] \sum_{r=1}^{K_i} \frac{-\triangledown f_i(w_t; x_{ir}) + \nu_{ir}}{bp_{i, t} K_i} \nonumber
    \\ + &\triangledown F(w_t)] \rangle \nonumber
    \\ &=\langle \triangledown F(w_t), \eta_t  \sum_{i, r}^{n, K_i} \text{Pr}[i \in \mathcal{Z}_{m, t}]E_{t}[\frac{-\triangledown f_i(w_t; x_{ir}) + \nu_{ir} + \triangledown f_i(w_t)}{bp_{i, t}K_i}]\rangle \nonumber
    \\ & \overset{(1)}{\leq} \langle  \sqrt{\eta_t} \triangledown F(w_t), \sum_{i, r}^{n, K_i} \frac{b p_{i, t} }{n} E_{t} [\frac{-\triangledown f_i(w_t; x_{ir}) + \nu_{ir} + \triangledown f_i(w_t)}{\frac{K_{min} b p_{i, t}}{\sqrt{\eta_t}}}] \rangle \nonumber
    \\ & = \langle \sqrt{\eta_t} \triangledown F(w_t), \sum_{i, r}^{n, K_i}\frac{\sqrt{\eta_t}}{nK_{min}} E_{t} [-\triangledown f_i(w_t; x_{ir}) + \nu_{ir} + \triangledown f_i(w_t)] \rangle
    \\ & \overset{(2)}{\leq}  \frac{\eta_t}{2}||\triangledown F(w_t)||^2 + \frac{\eta_t}{2nK_{min}} \cdot 2K_{max}n(\sigma_L^2 + c_{\nu}^2)
    \\ & = \frac{\eta_t}{2}||\triangledown F(w_t)||^2 + \frac{\eta_t K_{max}}{K_{min}} (\sigma_L^2 + c_{\nu}^2)
\end{align}
where (1) is a simple re-shuffling of constants and (2) uses the inequality $\langle a, b \rangle \leq \frac{1}{2}(||a||^2_2 + ||b||^2_2)$ and the analysis used in $B_1$ based on $E[||x_1 + x_2 + ... + x_n||^2] \leq n(E[||x_1||^2] + ... E[||x_n||^2)$.

Putting everything together, we get that,
\begin{align}
    \mathbb{E}_t&[F(w_{t+1})] \\ &\leq F(w_t) - \frac{\eta_t}{2} (1 - \sum_{i=1}^n \frac{L\eta_t(A^2 + 1)}{bnp_{i, t}}) ||\triangledown F(w_t)||^2_2 
    \\ & + \frac{(\eta_t K_{max}(\sigma^2_L + c_{\nu}^2))}{K_{min}}(1 + \sum_{i=1}^n \frac{L\eta_t}{bn p_{i, t}K_{min}}) 
    \\ & + \sum_{i=1}^n \frac{L\eta_t^2}{2bnp_{i, t}} (\sigma^2_G + d\sigma_0^2)
\end{align}

Re-arranging the terms, taking expectations for all the steps and summing from $0$ to $T-1$ steps (and canceling some terms), and dividing by $T$, we next get the following,
\begin{align}
    &\sum_{t=0}^{T-1} \mathbb{E}_t[||\triangledown F(w_t)||^2] (\frac{\eta}{2T} (1 - \sum_{i=1}^n \frac{L\eta_t^2 (A^2 + 1)}{bn p_{i, t}})) 
    \\ & \leq \frac{F(w_0) - F(w_T)}{T} + \sum_{t}^{T-1} \frac{(\eta_t K_{max}(\sigma^2_L + c_{\nu}^2))}{T K_{min}} 
    \\ & + \sum_{t, i}^{T-1,n} \frac{L\eta_t^2 K_{max} (\sigma^2_L + c_{\nu}^2)}{bnT p_{i, t}K_{min}^2} + \sum_{t, i}^{T-1, n} \frac{L\eta_t^2}{2bnTp_{i, t}} (\sigma^2_G + d\sigma_0^2)
    \\ & \overset{(1)}{\leq} \frac{F(w_0) - F(w_{*})}{T} + 
    \sum_{t}^{T-1} \frac{(\eta_b K_{max}(\sigma^2_L + c_{\nu}^2))}{K_{min} \sqrt{t+1} T}
    \\ & + \sum_{t, i}^{T-1,n} \frac{L\eta^2_b K_{max} (\sigma^2_L + c_{\nu}^2)}{bnT p_{i, t}K_{min}^2 (t+1)} + \sum_{t, i}^{T-1, n} \frac{L\eta_b^2}{2bnp_{i, t}(t+1)T} (\sigma^2_G + d\sigma_0^2)
    \\ & \overset{(2)}{\leq} \frac{F(w_0) - F(w_{*})}{T} + \frac{(2 \eta_b K_{max}(\sigma^2_L + c_{\nu}^2))}{K_{min}\sqrt{T}} 
    \\ & + \underbrace{\frac{L\eta_b^2}{bnT}[\frac{K_{max}(\sigma_L^2 + c_{\nu}^2)}{K_{min}} + \frac{\sigma^2_G + d\sigma^2_0}{2}]}_{C_1} \sum_{t, i} \frac{1}{p_{t, i}(t+1)}
\end{align}
Here, in (1) we assume that $\eta_{t} = \frac{\eta_{b}}{\sqrt{t+1}}$ where $\eta_b$ is the base learning rate. Also, we know that by definition, $F(w_{*}) \leq F(w_{T+1}) \rightarrow -F(w_{*}) \geq -F(w_{T+1})$. (2) follows from the idea that $\sum_{t=0}^{T-1} \frac{1}{\sqrt{t-1}} \leq \int_{1}^{T-1} \frac{1}{\sqrt{t-1}} = 2(\sqrt{T}) - 2 \leq 2\sqrt{T}$. 

We need to ensure that for an appropriate constant and all values of $t$, $\eta_b, \frac{L\eta_b^2 (A^2 + 1)}{bn} \sum_{i=1}^{n(t+1)} \frac{1}{p_{i, t}} \leq 1$. Note that the value is largest when $t = 1$, and so we essentially need,
\begin{align}\label{eqn:assumption-new}
     \sum_{i=1}^n \frac{1}{p_{i, t}} \leq \frac{b n}{L\eta_b^2 (A^2 + 1)}
\end{align}

Finally, we notice that to reduce the variance; we need a policy that follows equation~\ref{eqn:assumption-new}, $p_{i, t} \leq 1$ for all $i \in [n], t \in [0:T-1]$ combinations s.t.
\begin{align}
    \min_{p_{i, t} \leq 1, (\ref{eqn:assumption-new})} \sum_{t, i} \frac{C_1}{p_{i, t} (t+1)}
\end{align}

If we choose $p_{i, t} = \frac{1}{V_0}$ where $V_0 > 1$ is a constant, then we get the following value at round $t$,
\begin{align}
    &\frac{\eta_t}{2} (1 - \sum_i \frac{L\eta^2_t(A^2 + 1)}{bn p_{i, t}}) = \frac{\eta_t}{2} (1 -\frac{V_0 L\eta^2_t(A^2 + 1)}{b}) = \kappa_t
\end{align}
Taking summation over all values of $t$,
\begin{align}
    \sum_t \kappa_t &= \sum_t  \frac{\eta_t}{2} (1 -\frac{V_0 L\eta^2_t(A^2 + 1)}{b})
    \\ & = \eta_b (\sqrt{T} - 1) - \frac{V_0L\eta_b^2 log(T)(A^2+1)}{b} = \alpha_{\kappa} \sqrt{T}
\end{align}
where $\alpha_{\kappa} = \eta_b \frac{\sqrt{T} - 1}{\sqrt{T}} - \frac{V_0L\eta_b^2 log(T)(A^2+1)}{b\sqrt{T}}$.
Now, we can rewrite our expected variance function as,
\begin{align}
    &\min_t \mathbb{E}_t[||\triangledown F(w_t)||^2_2] \frac{1}{T} \sum_t \kappa_t
    \\ &= \frac{1}{T} \sum_{t} \min_t \mathbb{E}_t[||\triangledown F(w_t)||^2_2] \kappa_t
    \\ &\leq \frac{1}{T} \sum_{t} \mathbb{E}_t[||\triangledown F(w_t)||^2_2] \kappa_t
\end{align}

Substituting this back, our final result convergence result is,
\begin{align}
    &\min_t \mathbb{E}_t[||\triangledown F(w_t)||^2_2] 
    \\ &\leq \frac{1}{T} \sum_{t} \mathbb{E}_t[||\triangledown F(w_t)||^2_2] 
    \\ &= \mathcal{O}(\frac{\Delta F}{\sqrt{T}\alpha_{\kappa}} + \frac{\sigma^2_L + c_{\nu}^2}{\alpha_{\kappa}} + \frac{(\sigma^2_L + c_{\nu}^2 + \sigma^2_G + d\sigma_0^2) log(T)}{\sqrt{T} \alpha_{\kappa}})
    \\ & = \mathcal{O}(\frac{\Delta F}{\sqrt{T}} + \sigma^2_L + c_{\nu}^2 + \frac{(\sigma^2_L + c_{\nu}^2 + \sigma^2_G + d\sigma_0^2) log(T)}{\sqrt{T}})
\end{align}
Here, $\Delta F = F(w_0) - F(w_{*})$.
For this bound to be meaningful and assuming the bounded variances are reasonably small, the biggest challenge is the additional term of $dlog(T)\sigma^2_0$. We minimize variance only when $\sqrt{T} \approx dlog(T)$ for models with many parameters. After that, we get the bound for uniform sampling as,
\begin{align}
    \min_t \mathbb{E}_t[||\triangledown F(w_t)||^2_2] = \mathcal{O}(\frac{\Delta F}{\sqrt{T}} + \sigma_L^2 + c_{\nu}^2 + \frac{\sigma^2_G}{d} + \sigma_0^2)
\end{align}
This concludes the proof of the main theorem. The probabilities in the denominator should be large. But, with dynamic sampling some probabilities might be significantly small leading to large values for the variance. Therefore, we look at alternative ways of tackling the same problem.

We notice that the unbiasedness of the gradient can be maintained without probability scaling by assuming a weighted gradient output. 

\begin{lemma}\label{lemma:unscaled-unbiased-estimator}
    The private unscaled gradient estimate at round $t$, ($\tilde{\Delta}_t^u$) is an unbiased estimator of the aggregate clipped gradients, assuming that at each round $\sum_{i=1}^n p_{i, t} = 1$ and that the \textit{actual} gradient obtained is weighted. 
\end{lemma}
Let us first understand the expression for the unscaled $\tilde{\Delta}_t = \tilde{\Delta}_t^u$.
\begin{align}
    \mathbb{E}_t[\tilde{\Delta}_t^u] &= \mathbb{E}_t[\Delta_t^u + \bar{u}_t^u]
    \\ & \overset{(1)}{=} \mathbb{E}_t[\Delta_t]
    \\ & = \mathbb{E}_t[\frac{n}{b} \sum_{j \in \mathcal{Z}_{m. t}}{g}_j(w_t)]
    \\ &= \mathbb{E}_t[\frac{n}{b} \sum_{i = 1}^n \mathbbm{1}[i \in \mathcal{Z}_{m, t}] \bar{g}_i(w_t)]
    \\ &= \sum_{i = 1}^n \bar{g}_i(w_t) \frac{n}{b} \mathbb{E}_t[\mathbbm{1}[i \in \mathcal{Z}_{m, t}]]
    \\ & = \sum_{i = 1}^n \bar{g}_i(w_t) \frac{n}{b}
    \frac{1}{k} q_{i, t} p_{i, t}
    \\ & = \sum_{i = 1}^n \bar{g}_i(w_t) \frac{n}{b} \frac{m}{nk} p_{i, t}
    \\ & = \sum_{i = 1}^n \bar{g}_i(w_t)  p_{i, t}
\end{align}
Further, the expected batch size is $\sum_i \frac{q_{i, t}p_{i, t}}{k} =\sum_i \frac{mp_{i, t}}{kn} = \sum_i \frac{b}{n} p_{i, t} \leq \sum_i \frac{b}{n} p_{i, t} = b$ as $p_{i, t} \leq 1$. We believe this might be a better framework for analysis since this provides the following advantages:
\begin{itemize}
    \item Avoids having dynamic probabilities in the denominator.
    \item Sums up to 1 and provides weighted gradients to denote the weights intuitively.
    \item Maintains unbiased estimator, small batch size. 
\end{itemize}
The only minor inconvenience might be the $n$ factor in the numerator. We study the variance factor as part of our future work.

\begin{theorem}
    Algorithm~\ref{algorithm:LOCks} is $(\epsilon, \delta)$ differentially private.
\end{theorem}
\textbf{Proof:} We notice that in Algorithm~\ref{algorithm:LOCks}, the only private parameters are the raw gradients and their norms. Since, we bound each gradient by a clipped factor of $C$, we have bounded sensitivity. Normally, we would add noise with the variance $= \frac{Cc_0}{\epsilon_{step}}$ where $c_0 = 2\sqrt{\text{ln}(1.25/\delta_{step})}$ and $\epsilon_{step}, \delta_{step}$ are the per-step privacy budgets. With strong composition \cite{kairouz2015composition} we get,
\begin{align}
    \epsilon &= \epsilon_{step} \sqrt{2T \text{ln}(1/\delta_{step})} + T \frac{e^{\epsilon_{step}} - 1}{e^{\epsilon_{step}} + 1} 
    \\ & \approx \epsilon_{step} \sqrt{2T \text{ln}(1/\delta_{step})} + \frac{T\epsilon}{\epsilon + 1}
    \\ & \delta = (T+1)\delta_{step}
\end{align}

We add epsilons from the mechanisms with RDP \cite{mironov2017renyi}. However, the optimal value of $\alpha$ depends on the value of $\epsilon$ and $\delta$. Thus, given a target $\delta$, $\delta_{target}$ we have the following,
\begin{align}
    E_t[\epsilon] &\leq m\epsilon_s(\alpha_{*}) + b\epsilon_0(\alpha_{*}) \\
    & \delta = \delta_{target}
\end{align}
Here, $\alpha_{*}$ remains constant and can be replaced by the value that minimizes privacy costs. Further, the expectation is taken over the sampling randomness at step $t$, and the value of $\alpha_{*}$ is also dependent on the number of steps.

\section{Algorithms}

For the LOCks algorithm (Algorithm~\ref{algorithm:LOCks}), we provide details and explanations about the client-server architecture and how each update is processed.

In the first round, the server randomly picks a client w.p. $\frac{m}{n}$. These clients compute a private estimate of their last gradient norm. If the norm is negative, we clip to a minimal value $\approx 10^{-6}$. These norms are then converted to a probability distribution, and the ones with the highest probability are prioritized in the sampling. Thus, the probability is proportionate to the noisy gradient norm of the client. The clients picked to submit their private gradient to the server. The server aggregates this gradient and broadcasts the updated model to all the clients.

\begin{algorithm}
\caption{Generalized Optimal Client Sampling for Federated DP-SGD}\label{algorithm:LOCks}
\begin{algorithmic}[1]
\State \textbf{Input:} $n$ federated clients, privacy parameters $\epsilon, \delta$, number of iterations $T$, expected batch size $b$, maximum/minimum samples per client $K_{max}, K_{min}$, $\sigma_s, \sigma_0$ for gradient norms and gradient respectively identified by RDP analysis.
\State \textbf{Initialize:} model parameters $w_0 \in \mathbb{R}^p$, 
\For{t=0:T-1} \Comment{Total Iterations}
\State Get random sample $\mathcal{X}_{m, t}$ with sampling probability $\frac{m}{n}$
\For{$i \in \mathcal{X}_m$} \Comment{Computations at client $i$}
\State $g_i(w_{t}) \leftarrow  \sum_{j}^{K_i} \frac{\triangledown
f_i(w_{t}; x_{ij})}{{K_i}} \in \mathbb{R}^{p}$
\Comment{Client $i$ has $K_i$ samples}
\State $\bar{g}_i(w_{t}) \leftarrow \frac{g_i(w_{t})}{\max\{1, \frac{||g_i(w_{t})||_2}{C}\}}$ \Comment{Clip Gradients}
\State $v_i \sim \mathcal{N}(0, \sigma_s^2\mathbb{I}_{p})$
\State $\tilde{z}_i \leftarrow 
||\bar{g}_i(w_{t})||_2 + v_i$ \Comment{Private Gradient Norm}
\State Send $\tilde{z}_i$ to the server
\EndFor
\State $\tilde{z}_S \leftarrow \sum_{i \in \mathcal{X}_{t}} \tilde{z}_i$ \Comment{Server: Gradient norm sum over clients}
\State $\tilde{z}_S^{'} \leftarrow \min(\max(\tilde{z}_S, bC), \tilde{N}C)$ 
\For{$j \in \mathcal{X}_{t}$} \Comment{Computations at client $j$}
\State $p_{j, t} \leftarrow \min \{1, \frac{b\tilde{z}_j}{k\tilde{z}_S^{'}}\}$
\EndFor
\State Sample $\mathcal{Z}_{m, t}$ with the probability distribution derived from $q_j, j \in \mathcal{X}_t$ 
\For{$l \in \mathcal{Z}_{m, t}$} \Comment{Computations at client $l$}
\State $u_{l} \sim \mathcal{N}(0, \sigma^2_0)$
\EndFor
\State $\tilde{\Delta}_l \leftarrow \frac{1}{b} \sum_{l \in \mathcal{Z}_{m, t}} (\bar{g}_l (w_{t}) + u_{l})$ \Comment{Aggregation at server}
\State $w_{t+1} \leftarrow w_t -\eta_t\tilde{\Delta}_l$ \Comment{Server Update}
\State Return $w_{t+1}$ to all clients
\EndFor
\end{algorithmic}
\label{algo:main}
\end{algorithm}

\section{Experiments}

We provide preliminary experiments to motivate further investigation in this space. By uniformly sampling clients, we analyze a federated system. Here, the client only adds noise to the last gradient before sending it to the server. This provides user privacy and keeps the data on the device. For our analysis, we consider $100$ clients and assume that $20\%$ are uniformly randomly picked in each round. Each client runs $5$ local epochs between the server updates. The noise multiplier in each round is picked after searching for the best $\alpha$ value in RDP using Google's Tensorflow privacy library. We notice that for an $\epsilon = \infty$, we need a multiplier of $0$, for $\epsilon = 50$, we require a multiplier of $\approx 0.5$, and for $\epsilon = 10$, we need a multiplier of $1.32$. The smaller the multiplier, the lower the noise, but the higher the privacy loss leading to significant leaks.

Note that none of the provided $\epsilon$ values are satisfactory. We expect this since the gradient variance is proportionately higher if we use a smaller noise multiplier.

\begin{figure}[htb!]
\centering
\includegraphics[width=0.48\textwidth]{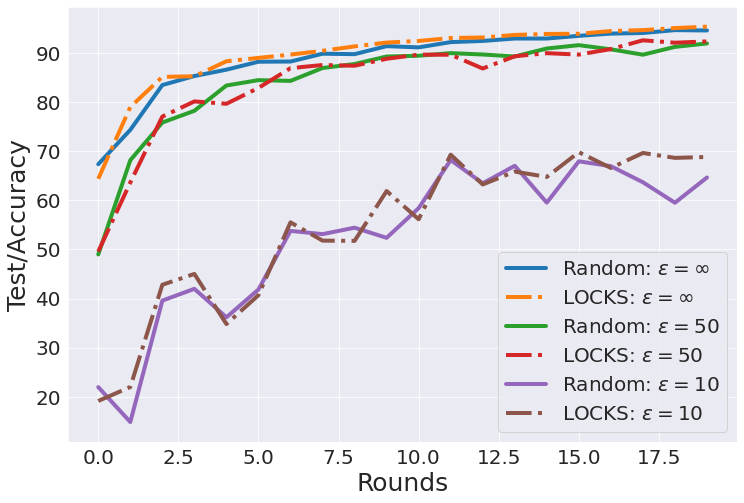}
\caption{Uniform Partial Client Sampling v/s LOCKS [MNIST]: Current sampling techniques do not perform well for simple datasets with user differential privacy. Further investigation is necessary to improve these mechanisms under noise.}
\label{fig:MNIST_alpha}
\end{figure}

We also empirically analyze the optimal client sampling techniques suggested in \cite{wei2022dpis} and \cite{li2019convergence} as modified in Algorithm~\ref{algorithm:LOCks}. In this scenario, we first randomly sample $30\%$ clients. We prioritize the clients with the highest private gradient norms in the second round. So, the probability of picking these clients is proportionate to their gradient size. By prioritizing the norm gradient, we see mild speed-ups in our convergence rate (empirically), as shown in Figure~\ref{fig:MNIST_alpha}. Though LOCKS generally performs better for high-noise / high-privacy cases, we still need to understand the underlying mechanism for its success.

We showcase our preliminary results for MNIST in Figure~\ref{fig:MNIST_alpha}. 

\section{Conclusions}

This work demonstrates the convergence of user differentially private and federated systems for non-convex objective functions with an adaptive learning rate and partial client participation. We provide a general two-pronged sampling framework that reduces the communication requirement for Federated systems per round. Our method maintains an unbiased gradient estimate while providing a flexible framework for further analysis. We also offer an alternative approach for analyzing the weighted sum of user gradients.
There are, however, significant gaps in the literature that suggest multiple open problems. 
\begin{itemize}
    \item \cite{wang2022unreasonable} shows that different gradients due to data heterogeneity do not impact empirical results. Therefore, we need to understand why gradient dissimilarity is important and how it impacts the convergence rate. 
    \item Currently, our model requires $d^2log(T)^2$ steps to reasonably converge. A high number of steps adds significant communication and privacy costs, and future work should tackle this issue. 
    \item Study how adaptive clipping and adaptive sampling affect convergence rates. 
    \item Demonstrate empirical results for complex datasets. 
    \item Alternative approaches for gradient similarity using similarity metrics like angles, cosine similarity, and correlation.
\end{itemize}

\bibliographystyle{ACM-Reference-Format}
\bibliography{main}

\appendix


\end{document}